\documentstyle[12pt]{article}
\begin{document}
\pagenumbering{roman}
\pagestyle{empty}
\begin{flushright}
Bartol Preprint BA-95-22\\
astro-ph/9506064
\end{flushright}
\begin{center}
{\Large \bf COBE OBSERVATIONS OF \\
THE MICROWAVE COUNTERPARTS OF \\
 GAMMA RAY BURSTS}
\end{center}
\vspace{0.5in}

\begin{center}
{R. K. Schaefer, S.Ali, M. Limon, and L. Piccirillo}
\end{center}
\vspace{0.1in}

\begin{center}
{\it Bartol Research Institute, \\
Univ. of Delaware, \\
Newark, DE 19716, USA}
\end{center}

\vspace{1.0in}
\begin{center}
Abstract:
\end{center}
\parbox{5.3in}{\small    We have used the data from the COBE satellite to
search for delayed
microwave emission (31 - 90 GHz) from Gamma Ray Bursts (GRBs).  The large
$7^\circ$ beam of COBE is well
matched to the large positional uncertainties in the GRB locations, although
it also means that fluxes from (point source) GRB objects will be diluted.  In
view of this we are doing a statistical search of the GRBs which occurred
during the currently released COBE DMR data (years 1990 and 1991), which
overlap $\sim 200$ GRBs recorded by GRO.  Here we concentrate on just the
top 10 GRBs (in peak counts/second).  We obtain the limits on the emission
by comparing the COBE fluxes before and after the GRB at the GRB location.
Since it is thought that the microwave emission should lag the GRB event, we
have searched the GRB position for emission in the few months
following the GRB occurrence. }

\vspace{0.25in}

\begin{center}
Contribution to 29th ESLAB symposium \\
{\it Towards the Source of Gamma Ray Bursts}
\end{center}

\newpage
\pagestyle{plain}
\pagenumbering{arabic}
\section{Microwave Observations by COBE}

   One of the most important techniques which can be used to identify the
sources of Gamma Ray Bursts (GRBs) is to find counterparts in other wavelength
regimes.  Here we focus on a search for delayed emission from GRBs in the
microwave region of the spectrum.  Such observations were made fortuitously by
the COBE (COsmic Background Explorer) satellite with the Differential Microwave
Radiometer (DMR) experiment.  COBE observes the full sky every 6 months at 3
frequencies (31, 53, and 90 GHz) with 12 radiometers.  (For a more complete
description of the COBE mission, see e.g., Boggess, et al, 1992.)  Some
models of GRBs (e.g., Paczynski and Rhoads, 1993; Meszaros and Rees, 1993;
Katz, 1994) predict microwave and radio emission which is delayed on time
scales of hours to months.  The best constraints on this type of delayed
emission in the microwave regime can be found from the COBE database.

    Unfortunately for this study, COBE was not designed for studying point
sources.  The radiometer horns have a FWHM of $7^\circ$ to purposely dilute the
effect of point sources and the radiometers
have rms noises of $\sim 23-59$ mK antenna temperature.  1 mK corresponds
to a point source emission of $\sim$ 1800 $(\nu/53$ GHz$)^2$ Janskys, so the
best use of COBE data is for studying delayed emission on very long time
scales where the noise can be reduced by integration to a useful level.  The
COBE  data has  also been used to put limits on prompt emission (see Bontekoe,
et al., these proceedings).  Another way to beat the radiometer noise is to
average over a  large ensemble of GRB.  Currently only 207 GRBs were recorded
by GRO during the period spanned by the publicly available COBE data.  This
number will be increased by a factor of $\sim 4$ when the remaining two years
of COBE data is released.  As a start towards this final goal we have analyzed
the top ten GRBs (rated by peak count rates) which have been recorded in the
released COBE data set.

\section{Background Subtraction and Analysis}

  COBE measures the difference in antenna temperature between the GRB position
and some reference point 60$^\circ$ away.  This angle defines a ring of
comparison points.  COBE orbits along the Earth's terminator, and as the Earth
moves in its orbit, the COBE viewing pattern sweeps through the sky (and hence
also the reference ring).  This reference ring is not uniform; it usually
slices through a portion of the galactic disk, which is a strong microwave
emitter.  This observing pattern therefore introduces a time dependent
background noise into the GRB observations as COBE sweeps through the different
portions of the reference ring.  The reference ring is swept through twice
per year, and the COBE orbit is slightly different in the biyearly
observations, so the simplest way to model this time dependent
reference signal is with two sine waves with a 1/2 and 1 (sidereal) year
periods. We fit the sine waves to the data taken at the GRB location for at
least one and a half years prior to the GRB event.  These fits are then
subtracted from the raw data.  The residuals should contain only the GRB source
emission.  We compare the residuals before and after the GRB to see if
they are different.

    We want to use these residuals to look for delayed emission from the GRB
source.  Unfortunately, we must have some model in mind to test the
data for goodness of fit.  The model cannot be very complicated for testing
against the limited sample here, because most of the GRB positions are only
observed part of the year, and the time interval between the GRB and the end of
the COBE data is different for each burst.  When the full data set is analysed,
we can test more sophisticated models, but for now we model the microwave GRB
signal as a simple increase in emission coincident with the GRB event and which
remains at a constant level thereafter.

\section{Results}

   All 10 of our selected GRBs are consistent with no change in emission after
the GRB.  We present our limits on the amount of emission after the event (in
Jy) in Table 1.  To insure that our background subtraction procedure is not
affecting our results, we have done two things.  1) We fit sine waves
on data only before the GRB and then subtract only the extrapolation from the
raw data after the GRB.  (This also reduces the accuracy of the fit slightly,
because it reduces the number of periods spanned by the data.)  This does
not significantly affect our results.  2) We also break up the residuals data
before the GRB into subsections and show that the pieces are also consistent
with zero emission.

\begin{table}[th]
\caption[]{95\% Confidence Upper Limits on the Change in Emission Level
Following the GRBs (integrated until the end of the COBE data--1/1/92). }
\begin{center}
\begin{tabular}{lllll}\hline
Rank  &  GRB & 31 Ghz U.L. &  53 GHz U.L. & 90GHz U.L. \\
      &      & (Jy)        & (Jy)         & (Jy)       \\\hline
1 & 910503  & 210 & 320 & 1900 \\
2 & 910601  & 120 & 240 & 720 \\
3 & 911118  & 736 & 643 & 1140\\
4 & 910609  & 290 & 240 &  830\\
5 & 910522  & 80  & 400 &  760\\
6 & 910627  & 330 & 360 &  540\\
7 & 910626  & 330 & 320 &  1270\\
8 & 911109  & 470 & 712 &  980\\
9 & 910421  & 350 & 270 &  910 \\
10 & 911104 & 580 & 670 &  2180\\ \hline
Likelihood & &  &  &  \\
Ratio &All 10 & 206 & 415 & 620 \\
\hline
\end{tabular}
\end{center}
\end{table}

   It would be useful to set fluence limits on these GRBs, but it is difficult
to make sensible limits when there are gaps the size of months in the GRB
coverage.  However, for two of the top ten (910522 and 910609) we have almost
``continuous" coverage (more than 200 1/2 second observations every day in the
first three months following the GRB, and only one gap of 2 days in the
remaining time because COBE was doing some internal calibration checks.  For
those two GRBs we calculate the fluence limits using the following formula
for microwave Fluence $F_\mu$:
\begin{equation}
F_\mu \equiv  \delta t \sum_{i=1}^3 \nu_i \langle F(\nu_i) \rangle ,
\end{equation}
where $\nu_i=31$, 53 and 90 GHz, $\langle F(\nu_i) \rangle$ is the
average (background subtracted) flux at frequency $\nu_i$ after the GRB.
$\delta t$ is the length of time over which observations (used to find
$\langle F(\nu_i) \rangle$) were made.  The results for these
fluences are given in table 2.

\begin{table}[thb]
\caption[]{95\% Confidence upper limits on fluence $F_\mu$ for two
gamma ray bursters.  The second column indicates how many days are
used in the fluence calculation.}
\begin{center}
\begin{tabular}{lll}\hline
GRB & Number of Days&  $F_\mu$ U.L. (ergs/cm$^2$) \\ \hline
910609  & 205 &  $8 \times 10^{5}$ \\
910522  & 223 &  $6 \times 10^{5}$ \\
\hline
\end{tabular}
\end{center}
\end{table}

\section{Conclusions}

We have analyzed the top ten GRBs (rated by BATSE peak counts) for delayed
emission in the 31-90 GHZ region as observed by COBE.  We find rather high
limits for the amount of possible microwave emission from these ten using a
crude model of the microwave emission.  Our results for the upper limits
on flux are presented in Table 1 and the limits on fluence for two GRBs are
presented in Table 2.

   We have selected these GRBs for two reasons: 1) if there is any effect it is
likely that the effect will be greatest in the most violent GRBs, and 2) we
needed a small subset to examine in detail to prepare for a larger study of
GRBs.  We expect to have the complete set of 207 Gamma Ray bursts analysed
soon.  With the complete ensemble, we expect more stringent limits smaller by a
factor of $1/\sqrt{20}$ and more detailed models can be more realistically
tested.  The release of the final part of the COBE data will increase the
number of Gamma ray bursts for study by a factor of 4, and will include such
events as the ``superbowl" burst.

 {\bf Acknowledgements:} We want to thank A. Kogut, D. Leisawitz, and J.
Newmark for help in deciphering the COBE data.  We also thank NASA for its
support in a grant from the ADP program.

\section{References}

\par
\hspace{0.22in}Boggess, N. {\it et al.}, 1992, ApJ, 397, 420.

\par
Bontekoe, T., {\it et al.}, 1995, {\sl Ast. \& Sp. Sci.}, (in press).

\par
Katz, J.~L., 1994, ApJ, 432, L107.

\par
Meszaros, P. \& Rees, M.~J., 1993, ApJ, 418, L59.

\par
Paczynski, B. \& Rhoads, J.~E., 1993, ApJ, 418, L5.
\par

\end{document}